\documentclass[aps,prb,twocolumn,superscriptaddress]{revtex4-1}
\usepackage{color}
\usepackage{dcolumn}   
\usepackage{bm}        
\usepackage{amssymb}   
\usepackage{amsmath}
\usepackage{gensymb}   
\usepackage{textcomp}
\usepackage{bbm}
\usepackage{hyperref}
\usepackage{cleveref}
\usepackage{graphicx}
\usepackage{array}
\usepackage{braket}
\usepackage{multirow}
\usepackage{natbib}

\newcommand{\Einq}{\ensuremath{< E <}\,}

\begin{document}

\title{\texorpdfstring{Spin gaps in the ordered states of La\textsubscript{2}LiXO\textsubscript{6} (X = Ru, Os) and their relation to distortion of the cubic double perovskite structure in 4\textit{d}\textsuperscript{3} and 5\textit{d}\textsuperscript{3} magnets.}{Spin gaps in the ordered states of La2LiXO6 (X = Ru, Os) and their relation to distortion of the cubic double perovskite structure in 4d3 and 5d3 magnets.}}

\author{D. D. Maharaj}
\email{maharadd@mcmaster.ca}
\affiliation{Department of Physics and Astronomy, McMaster University, Hamilton, ON L8S 4M1 Canada}

\author{G. Sala}
\affiliation{Department of Physics and Astronomy, McMaster University, Hamilton, ON L8S 4M1 Canada}
\affiliation{Neutron Scattering Division, Oak Ridge National Laboratory, Oak Ridge, Tennessee 37831, USA}

\author{C. A. Marjerrison}
\affiliation{Department of Physics and Astronomy, McMaster University, Hamilton, ON L8S 4M1 Canada}

\author{M. B. Stone}
\affiliation{Neutron Scattering Division, Oak Ridge National Laboratory, Oak Ridge, Tennessee 37831, USA}

\author{J. E. Greedan}
\affiliation{Department of Chemistry and Chemical Biology, McMaster University, ON, L8S 4M1, Canada}
\affiliation{Brockhouse Institute for Materials Research, McMaster University, Hamilton, ON L8S 4M1 Canada}

\author{B. D. Gaulin}
\affiliation{Department of Physics and Astronomy, McMaster University, Hamilton, ON L8S 4M1 Canada}
\affiliation{Brockhouse Institute for Materials Research, McMaster University, Hamilton, ON L8S 4M1 Canada}
\affiliation{Canadian Institute for Advanced Research, 661 University Ave., Toronto, ON M5G 1M1 Canada}

\date{\today}

\begin{abstract}
Time-of-flight inelastic neutron scattering measurements have been carried out on polycrystalline samples of the 4$d^3$ and 5$d^3$ double pervoskite antiferromagnets La$_2$LiRuO$_6$ and La$_2$LiOsO$_6$.  These reveal the development of an inelastic spin gap in La$_2$LiRuO$_6$ and La$_2$LiOsO$_6$ of $\sim$ 1.8(8) meV and 6(1) meV, below their respective ordering temperatures, $T_N$, $\sim$ 23.8 K and 30 K.  The bandwidth of the spin excitations is shown to be $\sim$ 5.7(9) to 12(1) meV, respectively, at low temperatures. Spin gaps are surprising in such magnets as the t$_{2g}$ levels of Ru$^{5+}$ or Os$^{5+}$ are expected to be half-filled, resulting in an anticipated orbital singlet for both materials. We compare these results in monoclinic double perovskites La$_2$LiRuO$_6$ and La$_2$LiOsO$_6$ with those in cubic Ba$_2$YRuO$_6$ and  Ba$_2$YOsO$_6$, as well as with those in the more strongly monoclinic La$_2$NaRuO$_6$, La$_2$NaOsO$_6$, and Sr$_2$ScOsO$_6$, and model the inelastic magnetic scattering  with linear spin wave theory using minimal anisotropic exchange interactions. We discuss the possible role of the distortion of the face-centered cubic, double perovskite structure on the spin gap formation and geometric frustration in these materials, and show that $T_N$ scales with the top of the spin wave band in all members of these families that display long range order.
\end{abstract}


\maketitle

\section{Introduction}

Double perovskite antiferromagnets display a diverse set of quantum magnetic ground states due to the confluence of geometrical frustration and strong spin-orbit coupling, two topical trends in contemporary condensed matter physics\cite{KimBalentsReview}.  Their low temperature phase behavior has been the subject of much recent study - both experimental and theoretical in nature\cite{Carlo1,Carlo2,Kermarrec,Thompson1,Thompson2,Aczel1,Aczel2,Taylor1,Taylor2,Dey,Das,Kanungo1,Li,Chen1,Chen2}. Double perovskites are characterized by a chemical formula of the form $A_2BB^{\prime}O_6$, where the \textit{B} and $B^{\prime}$ ions reside on octahedral sites, and form two interpenetrating face-centred cubic (FCC) lattices, provided that the overall structure is cubic. This is schematically shown in Fig. 1 a). If only one of the $B$ or $B^{\prime}$ sites is magnetic, such a sublattice forms a single magnetic FCC lattice and in this configuration the magnetic moments decorate a network of edge-sharing tetrahedra as seen in Fig. 1 b). This generates one of the canonical architectures supporting geometrical frustration in three dimensions\cite{Lacroix}.

These materials are such a rich platform for the study of quantum magnetism as the double perovskite structure is very flexible, and many magnetic and non-magnetic ions can occupy the $B$ and $B^{\prime}$ sublattices. The overall crystal symmetry can be lower than cubic\cite{Anderson} and, independently, the $B$ and $B^{\prime}$ sublattices can mix at some low ($\sim 5\%$) level. Monoclinic symmetries typically arise due to correlated rotations of $BO_6$ and $B^{\prime}O_6$ octahedra, as is shown in Fig. 1 c). Both $B$ - $B^{\prime}$ site mixing and distortions to structures with symmetry lower than cubic are controlled by the charge and ionic size difference between the $B$ and $B^{\prime}$ ions within the $A_2BB^{\prime}O_6$ structure\cite{Anderson}. In this regard, the study of families of double perovskite systems can enable systematic investigations of magnetic materials where the size of the moment, and its quantum nature, as well as the role of spin-orbit coupling, can be varied systematically within, or between, the many families of these materials. Strong spin-orbit coupling in 5$d$ systems is already known to induce a Mott instability in 5$d^5$ iridate compounds, leading to an effective total angular momentum $J_{eff} = \frac{1}{2}$ state, distinct from the $S =$ 1/2 localized state of conventional Mott insulators\cite{Kim}, and is hence a route to novel quantum states of matter at low temperatures.

The Ba$_2$YB'O$_6$ family of double perovskites illustrates well the diversity of magnetic ground states that can be realized when the magnetic $B^{\prime}$ ion is occupied by different 4$d$ or 5$d$ transition metal ions. This family has been of particular recent interest as its structure remains cubic to low temperatures, enabling the realization of the perfect frustrated FCC magnetic sublattice.  The family is also well ordered chemically, with only low levels ($\sim 1 \%$) of $B$ - $B^{\prime}$ of site mixing observed\cite{Aharen1,Aharen2,Aharen3}.  Specific members of this family studied to date include Ba$_2$YMoO$_6$ (4$d^1$), which exhibits a gapped collective spin-singlet ground state at low temperatures\cite{Carlo1}, Ba$_2$YReO$_6$ (4$d^2$) which shows an anomalous spin glass state below T$\sim$ 35 K \cite{Thompson2} and Ba$_2$YRuO$_6$ (4$d^3$) and Ba$_2$YOsO$_6$ (5$d^3$), which both form the same type I antiferromagnetic structure below $T_N$ $\sim$ 45 K and 69 K, respectively\cite{Carlo2,Kermarrec}. The structural and thermodynamic properties of Ba$_2$YIrO$_6$ (5$d^4$) have also been studied and it has been found to remain paramagnetic down to 0.4 K\cite{Dey}.

Here we consider the 4$d^3$ and 5$d^3$ double perovskites, La$_2$LiRuO$_6$ and  La$_2$LiOsO$_6$, with weak monoclinic distortions, and compare these new results to those previously obtained on the cubic $d^3$ systems, Ba$_2$YRuO$_6$ (4$d^3$) and Ba$_2$YOsO$_6$ (5$d^3$) as well as the more strongly monoclinic systems Sr$_2$ScOsO$_6$ and La$_2$NaRuO$_6$. La$_2$LiRuO$_6$ and  La$_2$LiOsO$_6$ display monoclinic angles $\beta = 90.020(5) \degree$ and $\beta$ = 90.147(1) $\degree$, respectively \cite{Thompson1,Battle}, which are small departures from $\beta = 90 \degree$ characterizing cubic symmetry. Like Ba$_2$YRuO$_6$ (4$d^3$) and Ba$_2$YOsO$_6$ (5$d^3$), La$_2$LiRuO$_6$ and La$_2$LiOsO$_6$ are expected to have the same half-filling of their $t_{2g}$ and thus display the same orbital singlet. La$_2$LiRuO$_6$ and  La$_2$LiOsO$_6$ are also expected to differ from each other, primarily, through the strength of the spin-orbit coupling displayed by 4$d^3$ compared to 5$d^3$. Earlier inelastic neutron scattering measurements (INS) on cubic Ba$_2$YRuO$_6$ and Ba$_2$YOsO$_6$ revealed spin gaps within their ordered states of $\sim$ 5 meV and 17 meV, respectively\cite{Carlo2,Kermarrec} with the 5$d^3$ gap being $\sim$ 3 times larger than that in the 4$d^3$ system\cite{Kermarrec}. The ratio of these spin gaps were found to be equal to the ratio of the free ion spin-orbit coupling factors, $\lambda$, for Os$^{5+}$ and Ru$^{5+}$, making a compelling argument that spin-orbit coupling stabilizes the $d^3$ spin gaps.

We have carried out a series of inelastic neutron scattering measurements on 4$d^3$ La$_2$LiRuO$_6$ and 5$d^3$ La$_2$LiOsO$_6$ in polycrystalline form.  The form of the inelastic magnetic scattering above and below their respective $T_N$s is seen to be qualitatively similar to that observed in Ba$_2$YRuO$_6$ and Ba$_2$YOsO$_6$, in that spin gaps develop coincident with $T_N$, and again scale roughly in proportion to expectations from atomic spin-orbit coupling factors.  We can quantitatively account for the ground state spin excitation spectra using classical linear spin wave theory and a minimal microscopic spin Hamiltonian involving near-neighbor anisotropic exchange.  We see that we get a very good description of the spin excitation spectra  from all four $d^3$ double perovskite materials, that is for the new inelastic scattering data from monoclinic La$_2$LiRuO$_6$ and La$_2$LiOsO$_6$, and from our earlier data on cubic Ba$_2$YRuO$_6$ and Ba$_2$YOsO$_6$.  This then allows us to make systematic comparisons between the microscopic spin Hamiltonian parameters so estimated in these systems, and to formulate an understanding of how the ordering temperatures, $T_N$, are related to each other and to their Hamiltonians.


\begin{figure}[h]
\includegraphics[width=8.6cm]{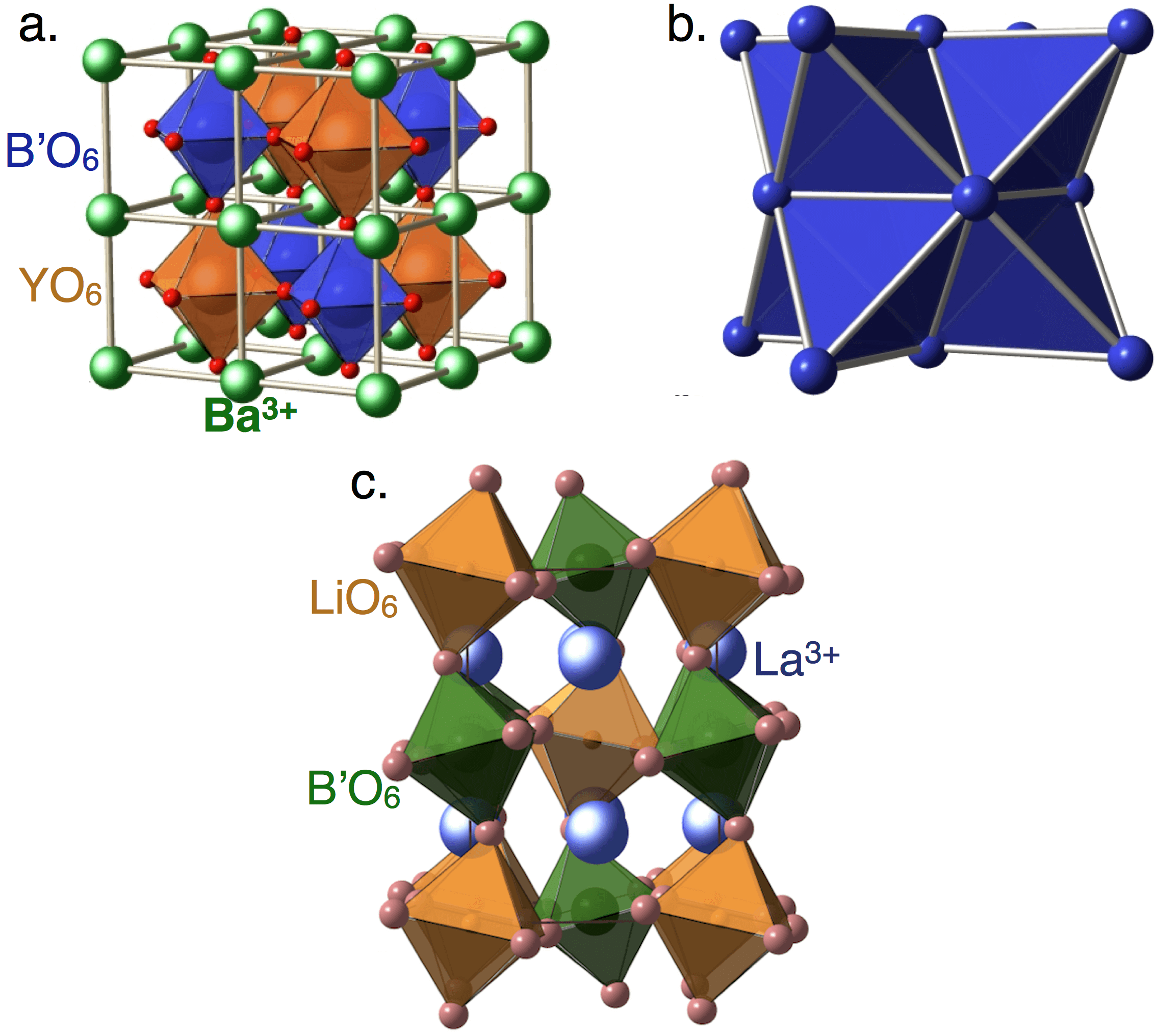}%
\caption{\label{LaLiXCS}The face-centred cubic (FCC) double perovskite structure exhibited by Ba$_2$YXO$_6$ (X = Ru, Os) and the lower symmetry structure of La$_2$LiXO$_6$ are shown in panels a) and c) respectively. In each structure the non-magnetic and magnetic B and B$^\prime$ ions co-ordinate with six oxygen atoms, forming a lattice of interpenetrating octahedra. The Ba$^{3+}$ and La$^{3+}$ atoms are distributed within this network. Panel b) shows the frustrated FCC network of edge-sharing tetrahedra of the B$^\prime$ site magnetic moments that are generated in the high symmetry Ba$_2$YXO$_6$ (X = Ru, Os) cubic structure.}
\end{figure}

\section{Experiment Details}

Time-of-flight INS measurements were performed using the direct geometry chopper spectrometer SEQUOIA, BL-17, located at the Spallation Neutron Source (SNS) of Oak Ridge National Laboratory~\cite{Sequoia}. Powder samples weighing 10 grams of each of La$_2$LiRuO$_6$ and  La$_2$LiOsO$_6$ were packed in aluminium foil and placed in identical aluminium annular cans, $\sim$ 3 cm in diameter. The two sample cans, as well as an empty sample can (used to obtain background measurements) were sealed in a glove box containing a He atmosphere to improve thermalization of the samples at low temperatures. The three cans were loaded on a three-sample carousel mounted to a closed-cycle refrigerator which produced a base temperature of 7 K.

Inelastic neutron scattering (INS) measurements were carried out on each sample using incident energies of  $E_i =$ 40 meV and 11 meV, which were both selected with chopper settings of $T_0 =$ 120 Hz and $FC_2 =$ 60 Hz, respectively. The elastic energy resolution associated with these INS measurements is $\sim$ 2$\%$ of E$_i$, giving elastic energy resolutions of $\sim$ 0.8 and 0.22 meV for $E_i =$ 40 meV and 11 meV, respectively.  These measurements were performed at a variety of temperatures above and below the respective N\'{e}el temperatures of La$_2$LiOsO$_6$ ($T_N$ = 30 K) and La$_2$LiRuO$_6$ ($T_N$ = 23.8 K). The data sets were reduced using Mantid\cite{mantid} and analyzed using neutron scattering software DAVE\cite{dave}. 

\section{Neutron Scattering Results and Calculations}

\subsection{Experiment Results and Analysis}

Representative plots of the neutron scattering intensity as a function of energy transfer, $\hbar\omega$, and wavevector transfer, $|Q|$, appropriate to the powder samples of La$_2$LiRuO$_6$ and La$_2$LiOsO$_6$ are shown in Fig. 2 and 3, respectively, for several temperatures near and below $T_N$. Figure 2 shows data taken on La$_2$LiRuO$_6$ using E$_i$ = 11 meV incident neutrons, while Fig. 3 shows data taken on La$_2$LiOsO$_6$ using E$_i$ = 40 meV incident neutrons.  In both cases an empty sample can data set has been subtracted as a background.  For both materials it is clear that a spin gap begins to develop near $T_N$, where $T_N$ = 23.8 K for La$_2$LiRuO$_6$ and $T_N$ = 30 K for La$_2$LiOsO$_6$. The spin gaps are well developed by $\frac{2}{3} \times$ $T_N$ and fully formed by our base temperature of $T =$ 7 K. The intensity scale for Figs. 2 and 3 is chosen to highlight the relatively weak inelastic scattering.  The much stronger elastic scattering saturates the scale in both figures, but clear Bragg peaks are observed to develop below $T_N$ at the 100 and 110 Bragg positions, near 0.8 \AA$^{-1}$ and 1.15 \AA$^{-1}$, as reported in a separate account of the magnetic elastic scattering and structure\cite{Thompson1}. For convenience, we employ the pseudo-cubic reciprocal lattice vector notations here and in the remainder of the paper.

\begin{figure}[h]
\includegraphics[width=8.6cm]{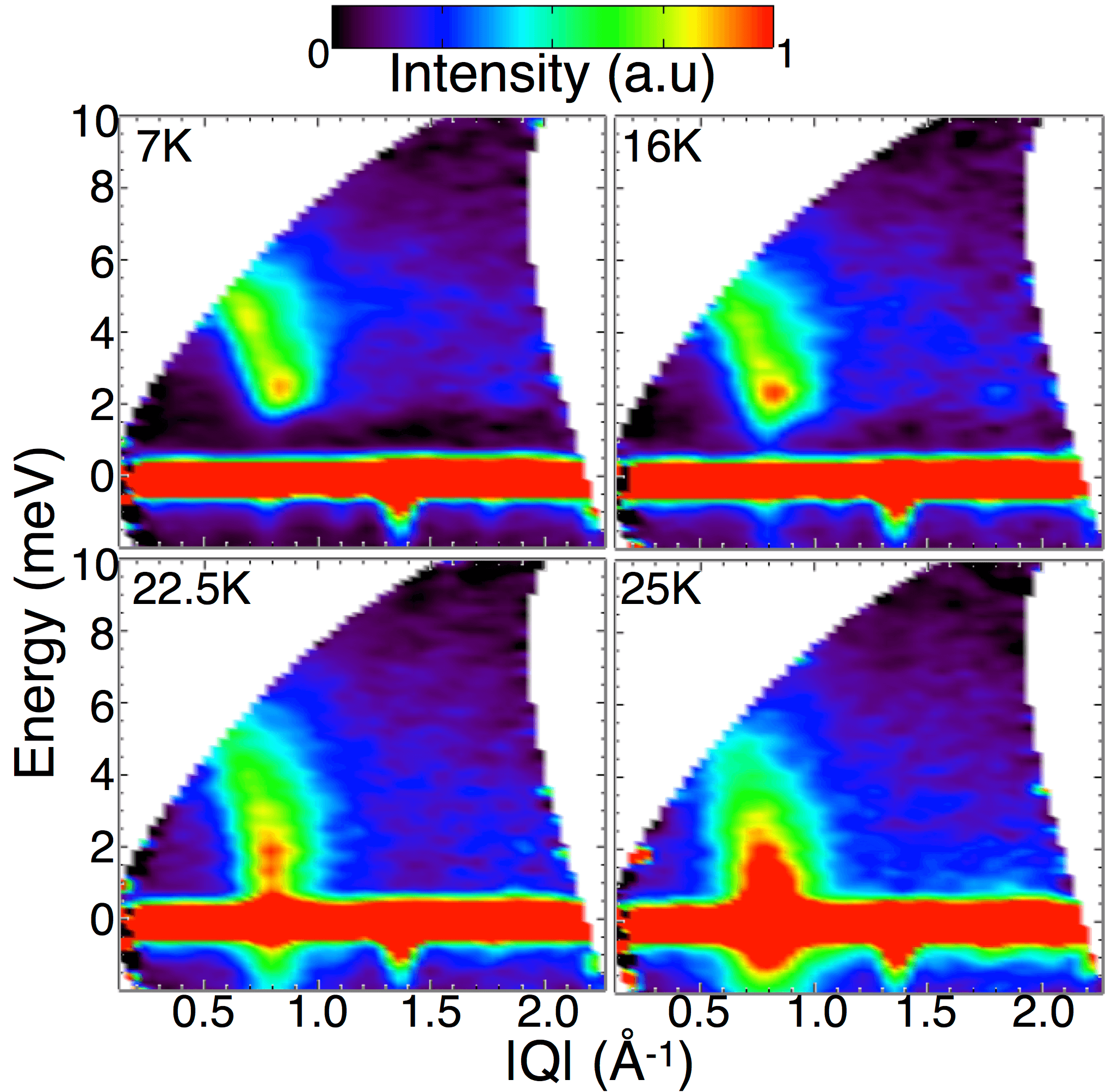}%
\caption{\label{RUCont}Contour plots showing the inelastic neutron scattering intensity as a function of energy transfer, $\hbar\omega$ and wavevector transfer, $|Q|$,  for La$_2$LiRuO$_6$ using $E_i =$ 11 meV neutrons. Above $T_N =$ 23.8 K there is an excess of quasi-elastic magnetic spectral weight centered near the (100) magnetic Bragg position at $|Q| =$ 0.8 \AA$^{-1}$. Below $T_N$, a spin gap develops and is fully formed by $T =$ 7 K. An empty sample cell data set at $T =$ 7 K has been subtracted from all displayed data sets.}
\end{figure}

\begin{figure}[h]
\includegraphics[width=8.6cm]{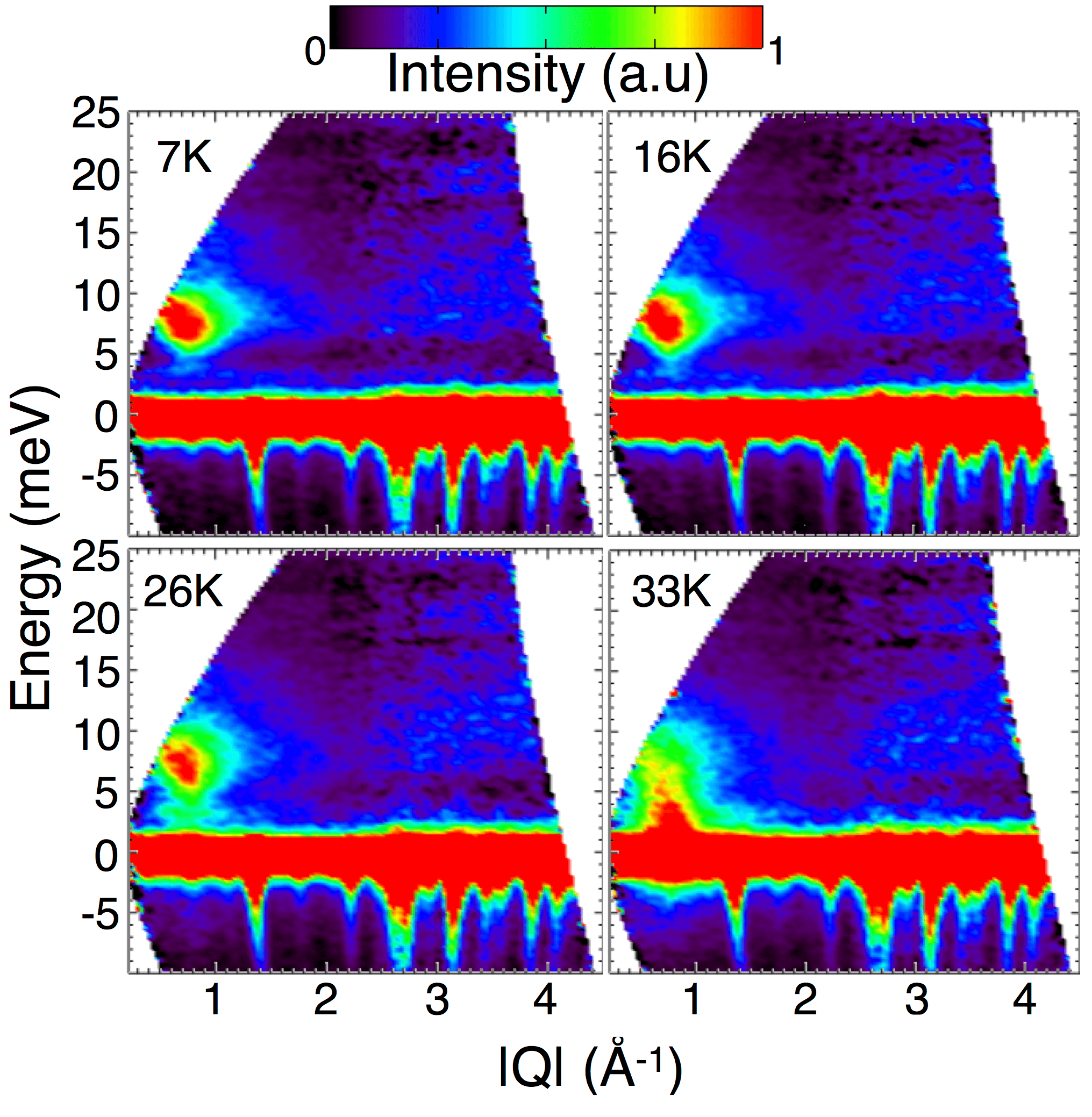}%
\caption{\label{OSCont}Contour plots showing the inelastic neutron scattering intensity as a function of energy transfer, $\hbar\omega$ and wavevector transfer, $|Q|$, are shown above for La$_2$LiOsO$_6$ using the $E_i =$ 40 meV neutron data set. Above $T_N =$ 30 K there is an excess of quasi-elastic magnetic spectral weight centered near the magnetic (100) Bragg position $|Q| =$ 0.8 \AA$^{-1}$. Below $T_N$, a spin gap develops and is fully formed by $T =$ 7 K. An empty sample cell data set at $T =$ 7 K has been subtracted from all displayed data sets.}
\end{figure}

\begin{figure}[h]
\includegraphics[width=8.6cm]{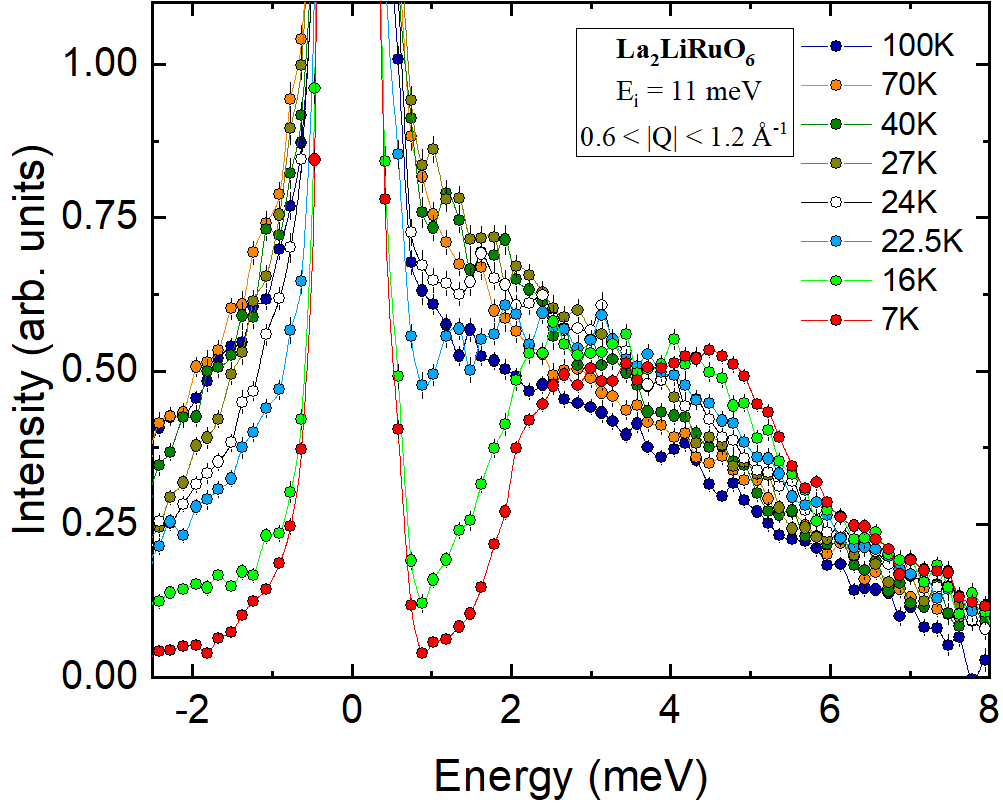}%
\caption{\label{RU_Inecut}$|Q|$ integrated ($|Q| =$ [0.6,1.2] \AA$^{-1}$) cuts of the inelastic neutron scattering data shown in Fig. 2, showing the $\sim$ 1.8 meV spin gap in La$_2$LiRuO$_6$. Above $T_N =$ 23.8 K, quasi-elastic magnetic spectral weight is observed.  At low temperatures, by $T = 7$ K, it is suppressed at energies low compared with the spin gap, and spectral weight shifts to higher energies where it is evident in a bimodal distribution of spin excitations with a total energy-bandwidth of $\sim$ 5.7 meV.}
\end{figure}

\begin{figure}[h]
\includegraphics[width=8.6cm]{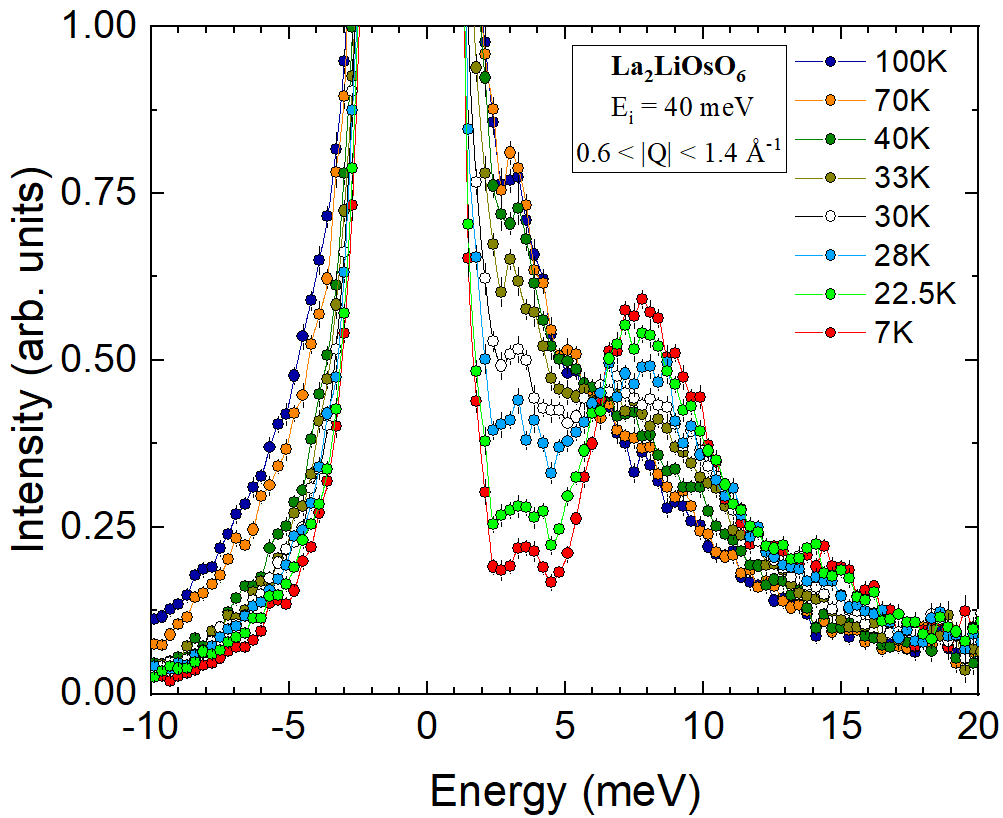}%
\caption{\label{OS_Inecut}$|Q|$ integrated ($|Q| =$ [0.6,1.4] \AA$^{-1}$) cuts showing the $\sim$ 6 meV spin gap in La$_2$LiOsO$_6$ at low temperatures. Above $T_N$ = 30 K, quasi-elastic magnetic spectral weight is observed.  Again the low energy scattering is suppressed below the spin gap below $T_N$.  At $T =$ 7 K a well developed spin gap is evident with spectral weight  transferred to energies above the gap.  The total energy-bandwidth of the spin excitations is $\sim$ 12 meV.}
\end{figure}

Detailed cuts through the two dimensional data sets in the $\hbar\omega$ - $|Q|$ maps of Figs. 2 and 3, are shown in Figs. 4 and 5, for La$_2$LiRuO$_6$ and La$_2$LiOsO$_6$, respectively.  These cuts are taken by integrating around the 100 and 110 positions in $|Q|$; 0.6 $<\,|Q|\,<$ 1.2 \AA$^{-1}$ for La$_2$LiRuO$_6$ with $E_i$ = 11 meV neutrons in Fig. 4, and 0.6 $<\,|Q|\,<$ 1.4 \AA$^{-1}$ for La$_2$LiOsO$_6$ with $E_i$ = 40 meV neutrons in Fig. 5. At our base temperature of $T =$ 7 K, we clearly identify a spin gap of 1.8(8) meV for La$_2$LiRuO$_6$ and 6(1) meV for La$_2$LiOsO$_6$. In both cases the spectral weight of the gapped magnetic scattering rises sharply from zero near the spin gap energy, and extends out with an energy-bandwidth of $\sim$ 5.9 meV for La$_2$LiRuO$_6$ and $\sim$ 12 meV for La$_2$LiOsO$_6$. The inelastic magnetic spectral weight in La$_2$LiRuO$_6$ appears to be bimodal, with the higher energy peak just below $\sim$ 5 meV, as can be seen in the low temperature data in Fig. 4.  For La$_2$LiOsO$_6$ in Fig. 5, one observes a extended high energy tail to the magnetic spectral weight above the spin gap at low temperatures. This is similar phenomenology to that displayed by the cubic double perovskites Ba$_2$YRuO$_6$ and Ba$_2$YOsO$_6$, where their low temperature magnetic spectral weight above their, larger, spin gaps, are a factor of 1.5 to 2 larger in bandwidth than those reported here observed in La$_2$LiRuO$_6$ and La$_2$LiOsO$_6$.

Figures 4 and 5 show in detail how the spin gap collapses in La$_2$LiRuO$_6$ and La$_2$LiOsO$_6$ and how the magnetic spectral weight fills in at low energies as the temperature moves towards and beyond their respective ordering temperatures.  This is shown more quantitatively in Figs. 6 and 7, where the temperature dependence of an integration of both the inelastic scattering, $S(|Q|,\omega)$, for energies above the spin gap, and the dynamic susceptibility, $\chi^{\prime\prime}$, for energies below the spin gap, is shown as a function of temperature for La$_2$LiRuO$_6$ and La$_2$LiOsO$_6$. 

$\chi^{\prime\prime}$ is related to the measured inelastic neutron scattered intensity by,

\begin{equation}\label{eq:Chi}
\Delta S(|Q|,\omega) = \frac{\chi^{\prime\prime}(|Q|,\hbar\omega)}{1-e^{-\hbar\omega/k_BT}}.
\end{equation}

\noindent Consideration of $\chi^{\prime\prime}$ allows the temperature dependence from detailed balance, contained in the Bose factor, $1-e^{-\hbar\omega/k_BT}$, to be removed, so that attention can focus on the physics of the system in question.  However this analysis depends on a good understanding of the background. Assuming that there is no inelastic scattering at energies well below the spin gap at low temperatures, we can use a low temperature data set as the background, $T =$ 7 K, and isolate $\chi^{\prime\prime}$ as a function of temperature.  The temperature dependence of this low energy $\chi^{\prime\prime}$ is shown in Fig. 6 a) and Fig. 7 a) for La$_2$LiRuO$_6$ and La$_2$LiOsO$_6$, respectively.   At energies above the spin gap, the inelastic magnetic scattering is never zero, and we look instead at the detailed temperature dependence of $S(|Q|,\omega)$ for La$_2$LiRuO$_6$ and La$_2$LiOsO$_6$ in Fig. 6 b) and 7 b), respectively.

Examination of Figs. 6 and 7 shows that $T_N$ occurs at the inflection points of either the growth of $\chi^{\prime\prime}$ below the spin gap, or the fall of $S(|Q|,\omega)$ above the spin gap, indicating that the formation of the spin gap is central to the magnetic phase transitions.  For both La$_2$LiRuO$_6$ and La$_2$LiOsO$_6$ materials $\chi^{\prime\prime}$ peaks at temperatures $\sim$ 10 - 20$\%$ above $T_N$, and then slowly decreases as temperature increases towards their respective Curie-Weiss temperatures, $\sim$ -204 K and -154 K, respectively.

The $|Q|$ dependence of $\chi^{\prime\prime}$ at low energies within the spin gap is shown for La$_2$LiRuO$_6$ and La$_2$LiOsO$_6$ in Fig. 8.  The energy integrations are performed over different ranges for La$_2$LiRuO$_6$ and La$_2$LiOsO$_6$ as the sizes of the spin gaps differ by a factor of $\sim$ 3.3.  The La$_2$LiRuO$_6$ data employs an energy integration from 0.9 meV \Einq 1.4 meV and Fig. 8 a) shows this $|Q|$ dependence as a function of temperature, for temperatures below and above $T_N$ = 23.8 K.  A very similar analysis is performed for La$_2$LiOsO$_6$ and the resulting $|Q|$ dependence of its low energy $\chi^{\prime\prime}$ is shown in Fig. 8 b) for temperatures below and above its $T_N$ = 30 K.  In this case the energy is integrated over a larger range, from 0.9 meV \Einq 4.0 meV.  

The trends in the $|Q|$ dependence of low energy $\chi^{\prime\prime}$ as a function of temperature are similar for the two materials.  The $|Q|$ values appropriate to the (100) and (110) ordering wavevectors are denoted with vertical red fiduciaries in both panels of Fig. 8.  One can see that $\chi^{\prime\prime}$ ($|Q|$, $\hbar\omega < \Delta$) is centred primarily on the (100) ordering wavevector, and this peak rises in intensity as the temperature approaches $T_N$.  The $|Q|$ dependence of $\chi^{\prime\prime}$ ($|Q|$, $\hbar\omega < \Delta$) for La$_2$LiOsO$_6$ is, however, clearly broader than that of La$_2$LiRuO$_6$.  This is likely a reflection of the different (1/2 1/2 0) magnetic ordering wavevector that  La$_2$LiOsO$_6$ displays, compared with the (100) type I antiferromagnetic ordering that La$_2$LiRuO$_6$ displays.  The relative structure factor for the (110) magnetic Bragg intensity, compared with the (100) magnetic Bragg intensity, is stronger for the (1/2 1/2 0) structure displayed by La$_2$LiOsO$_6$ than for the (100) structure displayed by La$_2$LiRuO$_6$.  A natural explanation for the increased breath in $|Q|$ for La$_2$LiOsO$_6$, is that the dynamic spectral weight is also relatively stronger at (110) and this extends the $|Q|$ dependence in $\chi^{\prime\prime}$ from (100) to (110), therefore out to larger $|Q|$s.  As will be discussed below in the context of linear spin wave theory applied to these systems, the observed magnetic scattering in both systems falls off anomalously quickly with $|Q|$, likely related to co-valency of the 4 and 5$d$-electrons, which implies a magnetic form factor corresponding to more extended d electron wavefunctions.  This effect also tends to concentrate the inelastic scattering shown in Fig. 8  to smaller $|Q|$ than would otherwise be the case.

\begin{figure}[htbp!]
\includegraphics[width=8.6cm]{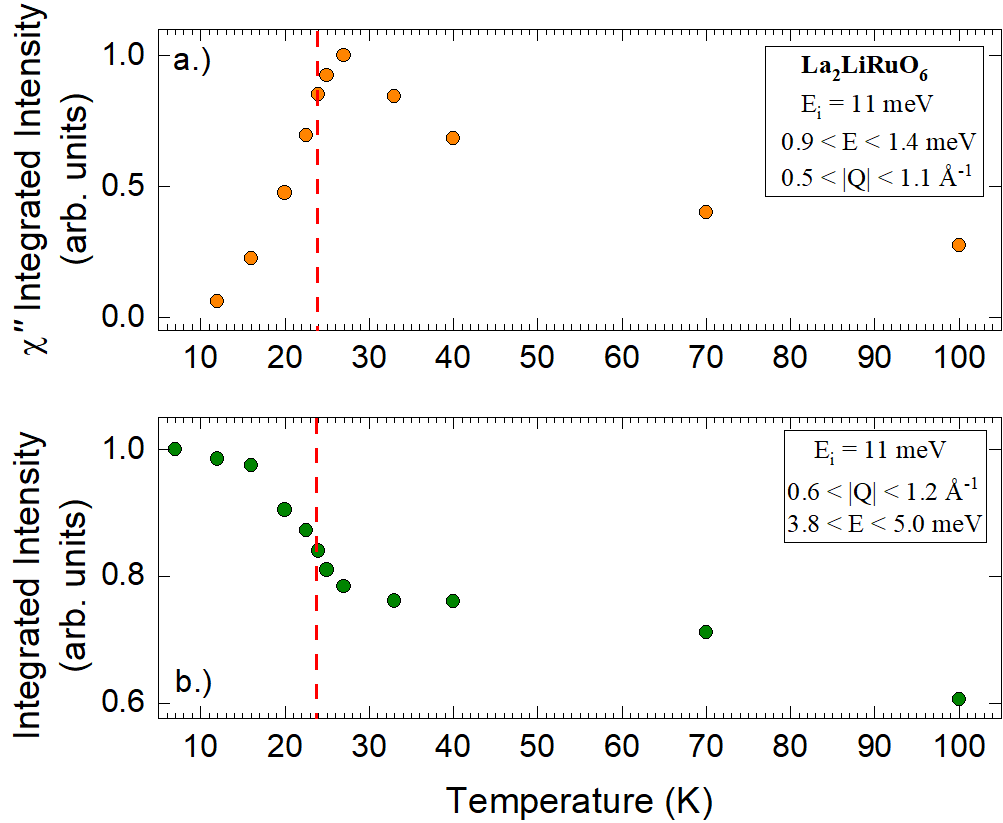}%
\caption{\label{RU_II}The temperature dependence of $\chi^{\prime\prime}$ for La$_2$LiRuO$_6$ is shown in panel a) where the integration of the intensity was performed in the range $|Q| =$ [0.5, 1.1] \AA$^{-1}$ and E = [0.9, 1.4] meV $< \Delta$.  Data taken at $T =$ 7 K has been used as a background.  This is derived from integrals of the data presented in Fig. 2.  A complementary plot of the scattered intensity obtained from integrating $|Q| =$ [0.6, 1.2] \AA$^{-1}$ and E = [3.8, 4.9] meV $> \Delta$ is shown in b). $T_N$ is shown as the vertical dashed line in both panels.}
\end{figure}

\begin{figure}[htbp!]
\includegraphics[width=8.6cm]{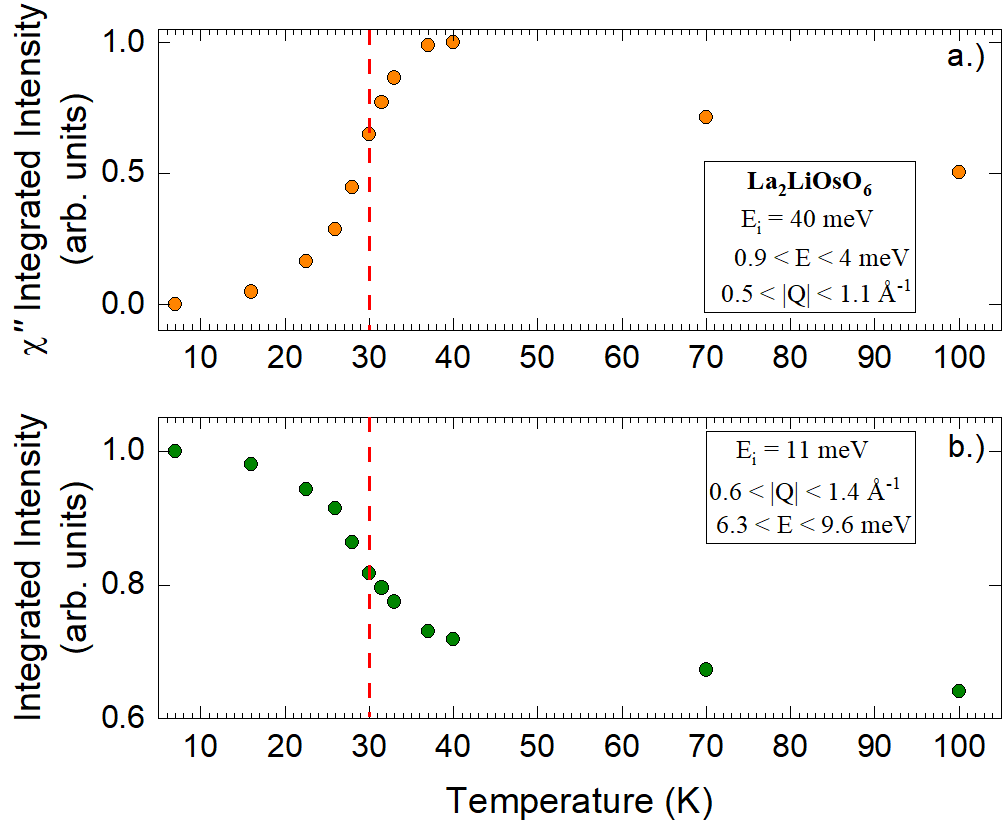}%
\caption{\label{OS_II}The temperature dependence of $\chi^{\prime\prime}$ for La$_2$LiOsO$_6$ is shown in a) where the integration of the intensity was performed in the range $|Q|$ = [0.5, 1.1] \AA$^{-1}$ and E = [0.9, 4] meV $<\,\Delta$.  Data taken at $T =$ 7 K has been used as a background. This is derived from integrals of the data presented in Fig. 3.  A complementary plot of the scattered intensity obtained by integrating $|Q| =$ [0.6, 1.4] \AA$^{-1}$ and E = [6.3, 9.6] meV $>\,\Delta$) is shown in b). $T_N$ is shown as the vertical dashed line in both panels.}
\end{figure}

\begin{figure}[htbp!]
\includegraphics[width=8.6cm]{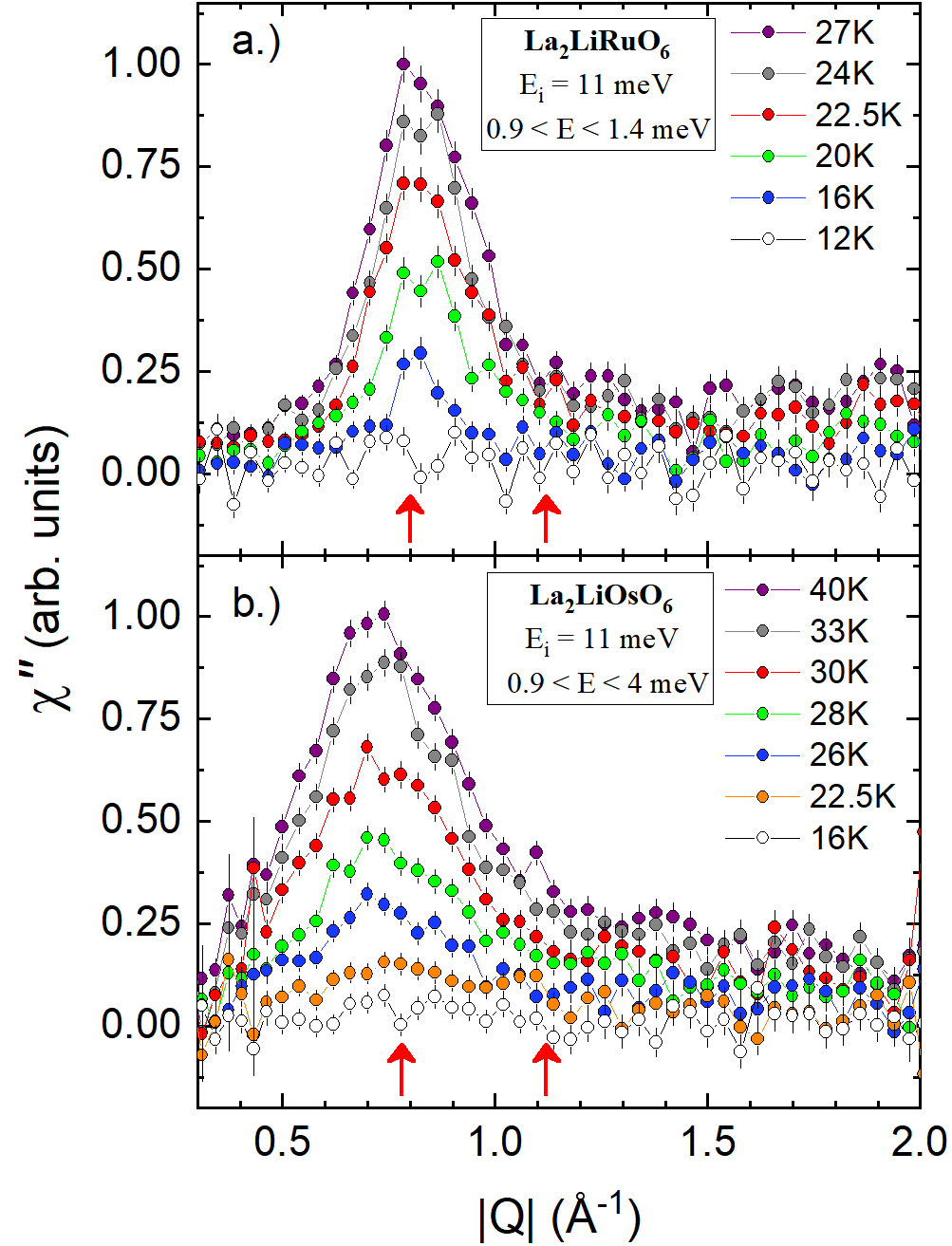}%
\caption{\label{RU_Chi} a) The temperature dependence of $\chi^{\prime\prime}$  for La$_2$LiRuO$_6$ is shown where integrations were performed with $|Q| =$ [0.5, 1.1] \AA$^{-1}$ and $E =$ [0.9, 1.4] meV $<\,\Delta$. Data taken at $T =$ 7 K has been used as a background.  This shows the $|Q|$ dependence of $\chi^{\prime\prime}$ ($|Q|$, $E\,<\,\Delta$) and how it evolves as a function of temperature for temperatures below and just above $T_N$. b) The temperature dependence of low energy cuts of $\chi^{\prime\prime}$ taken in the range $E =$ [0.9, 4] meV for La$_2$LiOsO$_6$ are shown. This shows the $|Q|$ dependence of $\chi^{\prime\prime}$ ($|Q|$, E $<\,\Delta$) and how it evolves as a function of temperature for temperatures below and just above $T_N$. In both panels, the red fiducial arrows indicate the positions of the (100) and (110) magnetic Bragg positions.}
\end{figure}

\subsection{Linear Spin Wave Theory Calculations{\label{SpinW}}}

Linear spin wave theory calculations were carried out in order to estimate the microscopic spin Hamiltonian for the double perovksite systems La$_2$LiRuO$_6$, La$_2$LiOsO$_6$, Ba$_2$YRuO$_6$ and Ba$_2$YOsO$_6$.  The calculations were performed using the SpinW software package.\cite{Toth}, and these were benchmarked against the ground state ($T =$ 7 K) inelastic neutron scattering data for La$_2$LiRuO$_6$ at $T =$ 7 K in Fig. 2, for La$_2$LiOsO$_6$ at $T =$ 7 K in Fig. 3, and using the corresponding data for Ba$_2$YRuO$_6$, Ba$_2$YOsO$_6$ and La$_2$NaRuO$_6$, taken from previous studies performed by Carlo \textit{et al.}\citep{Carlo2}, Kermarrec \textit{et al}\cite{Kermarrec} and Aczel \textit{et al.}, respectively.

The linear spin wave theory employed a minimal model for an anisotropic exchange Hamiltonian, which reproduces the type I antiferromagnetic ordered state displayed by La$_2$LiRuO$_6$, Ba$_2$YRuO$_6$, and Ba$_2$YOsO$_6$ at low temperatures, and which also produces a gap in the low energy magnetic inelastic spectrum.

\begin{eqnarray}
\mathcal{H} = -\Big(J_1 \sum_{NN} \mathbf{S}_i\mathbf{S}_j + K_1 \sum_{NN} S_{i,x}S_{j,x}\Big),
\label{eq:HamK}
\end{eqnarray}

\noindent where the \textit{J}$_1$ term represents an isotropic near-neighbor exchange interaction, related to the bandwidth of the spin excitations. The \textit{K}$_1$ term generates the spin gap in these systems and an additional coupling of a particular component of spin, hence anisotropic exchange. \textit{J}$_1$ and \textit{K}$_1$ are defined such that positive values are ferromagnetic and negative values are antiferromagnetic. Near-neighbor anisotropic exchange interactions were previously identified as the likely cause for the spin gap observed in these $d^3$ systems\cite{Kermarrec,Taylor2}, and a similar spin wave theory analysis was carried out by Taylor \textit{et al}\cite{Taylor2} to model the spin excitation spectrum of the related monoclinic $d^3$ double perovskite, Sr$_2$ScOsO$_6$. The spin wave theory calculations were performed equivalently for both the cubic and monoclinic systems despite the slight distortion in the monoclinic systems, La$_2$LiOsO$_6$ and La$_2$LiRuO$_6$.  These distortions are relatively weak and their inelastic neutron scattering spectra are remarkably similar to their cubic counterparts. The Os$^{5+}$ magnetic form factor, obtained from work of Kobayashi \textit{et al}\cite{Kobayashi}, was employed in our spin wave theory calculations. To our knowledge, the magnetic form factor for Ru$^{5+}$ is not reported in the literature and as such, the Os$^{5+}$ form factor was also used in the spin wave calculations for the Ru$^{5+}$ systems.

\begin{figure}[htbp!]
\includegraphics[width=8.6cm]{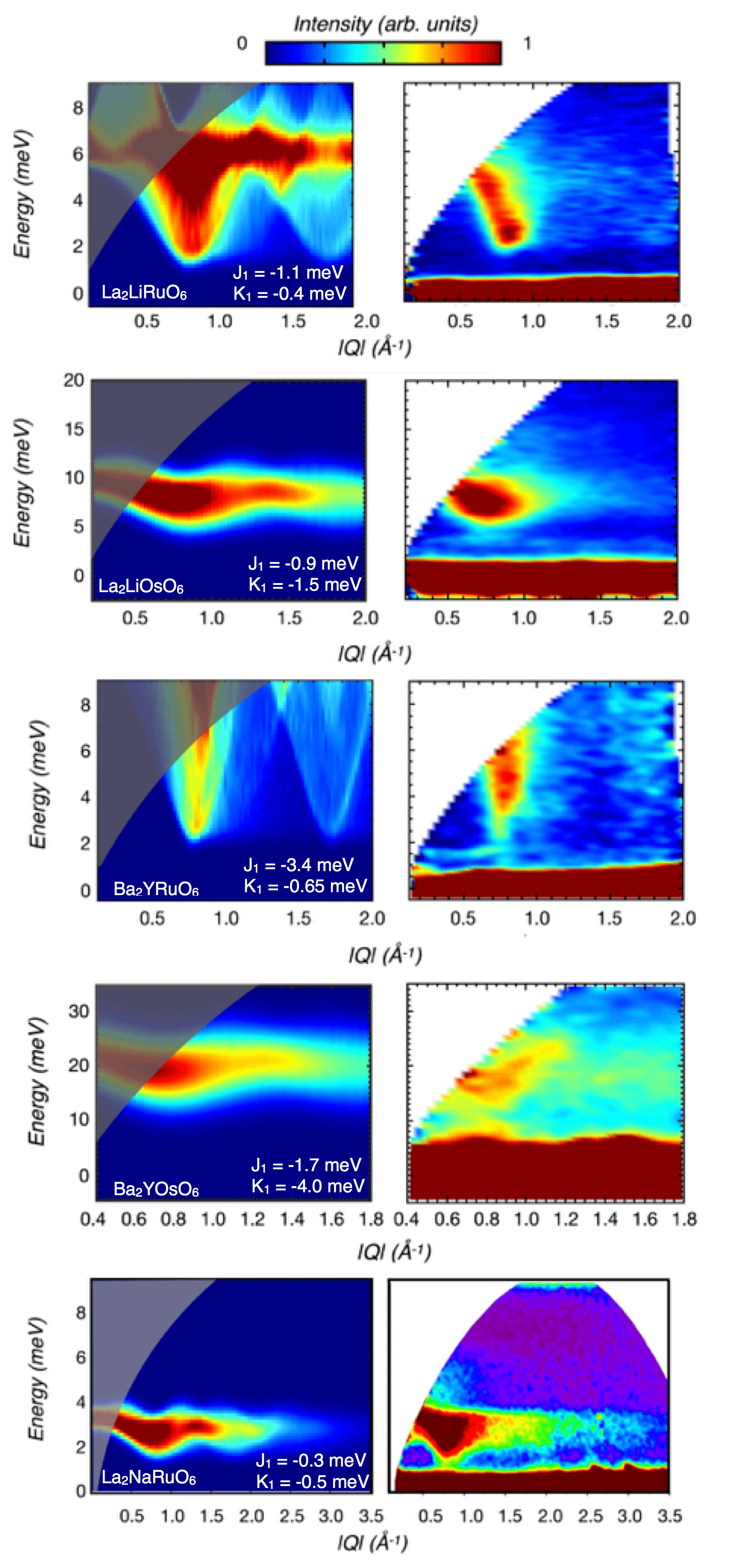}%
\caption{\label{fig:Compiled_Calcs}The calculated spin wave spectra, performed with SpinW for the double perovskites, La$_2$LiRuO$_6$, La$_2$LiOsO$_6$, Ba$_2$YRuO$_6$, Ba$_2$YOsO$_6$, and La$_2$NaRuO$_6$ are presented in the left panels and the corresponding experimentally obtained INS spectra obtained at base temperatures are provided on the right. All experimental data sets are background subtracted.  Appropriate Gaussian broadening of the calculated spin wave spectra was applied to each case, in order to account for experimental resolution.}
\end{figure}

The spectra obtained from calculations using equation 2 are shown in \cref{fig:Compiled_Calcs} and the resulting fit parameters are presented in \cref{tab:fit_parameters}. The parameters provide good phenomenological descriptions of the observed spectra at low temperatures. The primary difference between the observed spectra and calculated spectra is that the intensity of the magnetic excitations drops off more rapidly as a function of $|Q|$ in the former case. This can be explained in terms of metal-ligand covalency effects between the $d$ orbitals of the B$^{\prime}$ ions and the $p$ orbitals of the neighboring $O^{2-}$ ions\cite{Kermarrec,Taylor2,Granado,Aczel2}.  Thus the magnetic form factor for both Os$^{5+}$ and Ru$^{5+}$ should reflect more extended $d$-electron wavefunctions, and thus drop off more sharply with $|Q|$.

The spin wave theory calculations were carried out, adjusting the two parameters in the spin Hamiltonian, $J_1$ and $K_1$, using the inelastic neutron scattering for La$_2$LiRuO$_6$, La$_2$LiOsO$_6$, Ba$_2$YRuO$_6$ and Ba$_2$YOsO$_6$ as benchmarks, until a good description of the data was achieved.  For comparison, literature data for La$_2$NaRuO$_6$ was also fit in this manner.  Our best efforts resulted in the comparison between theory and experiment shown in Fig. 9.  The resulting ``best fit" parameters for $J_1$ and $K_1$ are shown both in the appropriate panels of Fig. 9 and are listed in Table 1 for all five double perovskites, as well as for Sr$_2$ScOsO$_6$, using the literature results from Taylor et al \cite{Taylor2}.  Comparing now between the Ba$_2$YRu/OsO$_6$ and La$_2$LiRu/OsO$_6$ families,  it is clear that all the energy scales are higher in cubic Ba$_2$YRu/OsO$_6$ compared with La$_2$LiRu/OsO$_6$, consistent with the $\theta_{CW}$ values being higher in the cubic Ba$_2$YRu/OsO$_6$ family.  Looking across the $J_1$ and $K_1$ parameters for the six douple perovskites listed in Table I, we see that the anisotropic exchange values, $K_1$ are relatively consistent, $\sim$ -0.5 meV for the ruthenates, and from -1.5 meV to -4 meV for the osmates.  There is greater variation in the isotropic exchange parameter, $J_1$, ranging from -0.3 meV for La$_2$NaRuO$_6$ which displays an incommensurate magnetic structure to -4.4 meV for Sr$_2$ScOsO$_6$ which displays the highest temperature phase transition, $T_N$ = 92 K, to a type I antiferromagnetic structure.  Where the comparison can be made, in the Ba$_2$YRu/OsO$_6$ and La$_2$LiRu/OsO$_6$ families, the anisotropic near-neighbor exchange, $K_1$, is much stronger in the osmate member of each family relative to the ruthenate member.  This latter effect is responsible for the much higher spin gaps in the osmate members of the families compared to the ruthenates, and is consistent with spin-orbit coupling being $\sim$ 3.2 times stronger in 5$d^3$ Os$^{5+}$ compared to 4$d^3$ Ru$^{5+}$ configurations.  

The inelastic neutron scattering spectra can also be effectively reproduced using a spin Hamiltonian with isotropic near-neighbor exchange and single-ion anisotropy, as

\begin{eqnarray}
\mathcal{H} = -\Big(J_1 \sum_{NN} \mathbf{S}_i\mathbf{S}_j + D \sum_{i} S^2_{i,x}\Big),
\label{eq:HamD}
\end{eqnarray}

The quality of the comparison between experiment and theory using this spin Hamiltonian is very similar to that using anisotropic exchange with Eq. 2, and hence this comparison is not reproduced here.  The best fits to the inelastic spectra for La$_2$LiRuO$_6$, La$_2$LiOsO$_6$, Ba$_2$YRuO$_6$ and Ba$_2$YOsO$_6$, as well as Sr$_2$ScOsO$_6$ and La$_2$NaRuO$_6$, are listed in Table 1, where the same \textit{J}$_1$ value is relevant to best fits with either single-ion anistropy (Eq. 2) or anisotropic exchange (Eq. 1).


\begin{table} 
\centering
\begin{ruledtabular}\begin{tabular}{cccccc}
System & $J_1$ (meV) & $K_1$ (meV) & $D$ (meV) & $\frac{K_1}{J_1}$ & $\frac{|D|}{J_1}$ \\
\hline
$\mathrm{Ba_2YRuO_6}$ & -3.4 & -0.65 & 1.3 & 0.2 & 0.4 \\
$\mathrm{Ba_2YOsO_6}$ & -1.7 & -4.0 & 8.0 & 2.4 & 4.7 \\
\hline
$\mathrm{La_2LiRuO_6}$ & -1.1 & -0.4 & 0.8 & 0.4 & 0.7 \\
$\mathrm{La_2LiOsO_6}$ & -0.9 &  -1.5 &  2.9 & 1.7 & 3.2 \\
\hline
$\mathrm{Sr_2ScOsO_6}$ & -4.4 & -3.8 & 7.5 & 0.9 & 1.7 \\
\hline
$\mathrm{La_2NaRuO_6}$ & -0.3 & -0.5 & 1.0 & 1.7 & 3.3 \\
\end{tabular}\end{ruledtabular}
\caption{\label{tab:fit_parameters}
Microscopic exchange parameters relevant to each double perovskite system resulting from SpinW fits to the experimental spectra are shown.  \textit{J}$_1$, \textit{K}$_1$ and \textit{D} represent isotropic nearest neighbor exchange, anisotropic nearest neighbor exchange and single-ion Ising-like anisotropy, respectively.}
\end{table}

\begin{figure}[htbp!]
\includegraphics[width=8.6cm]{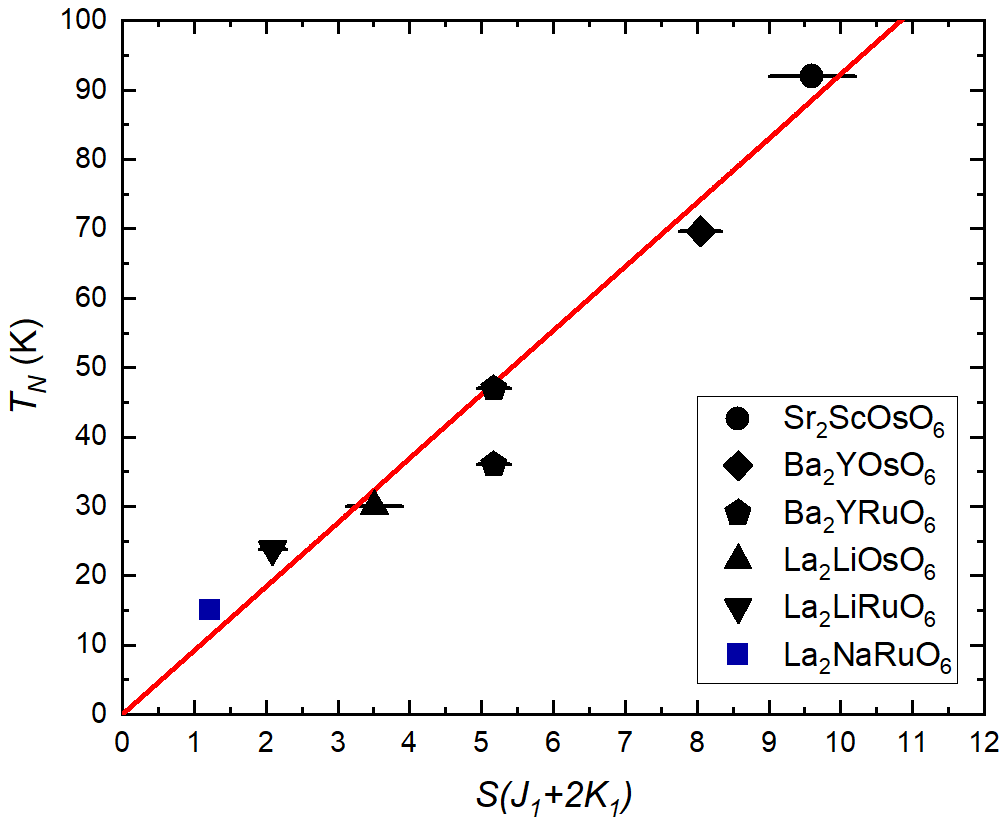}%
\caption{\label{fig:TN_vs_Exchange}The correlation between $S(J_1+2K_1)$ versus ordering temperature, $T_N$ is shown for La$_2$LiRuO$_6$, La$_2$LiOsO$_6$, Ba$_2$YRuO$_6$, Ba$_2$YOsO$_6$, Sr$_2$ScOsO$_6$ and La$_2$NaRuO$_6$. The $T_N$ for Ba$_2$YRuO$_6$ employed in this plot is the higher of the two transition temperatures, T$_N \sim$ 45 K, relevant for Ba$_2$YRuO$_6$. Please note that i.) the error bars associated with $S(J_1+2K_1)$ are due to $S$ and as such, we expect this to be a lower limit on the estimate of the error in this quantity. ii.) the data point corresponding to La$_2$NaRuO$_6$ is highlighted in blue to emphasize that a $k =$ (0,0,0) type I AF magnetic structure was utilized for simplicity in the SpinW calculation. In actuality, the material exhibits an unusual incommensurate magnetic structure as reported in original work by Aczel \textit{et al}\citep{Aczel2}.}
\end{figure}

\begin{figure}[htbp!]
\includegraphics[width=8.6cm]{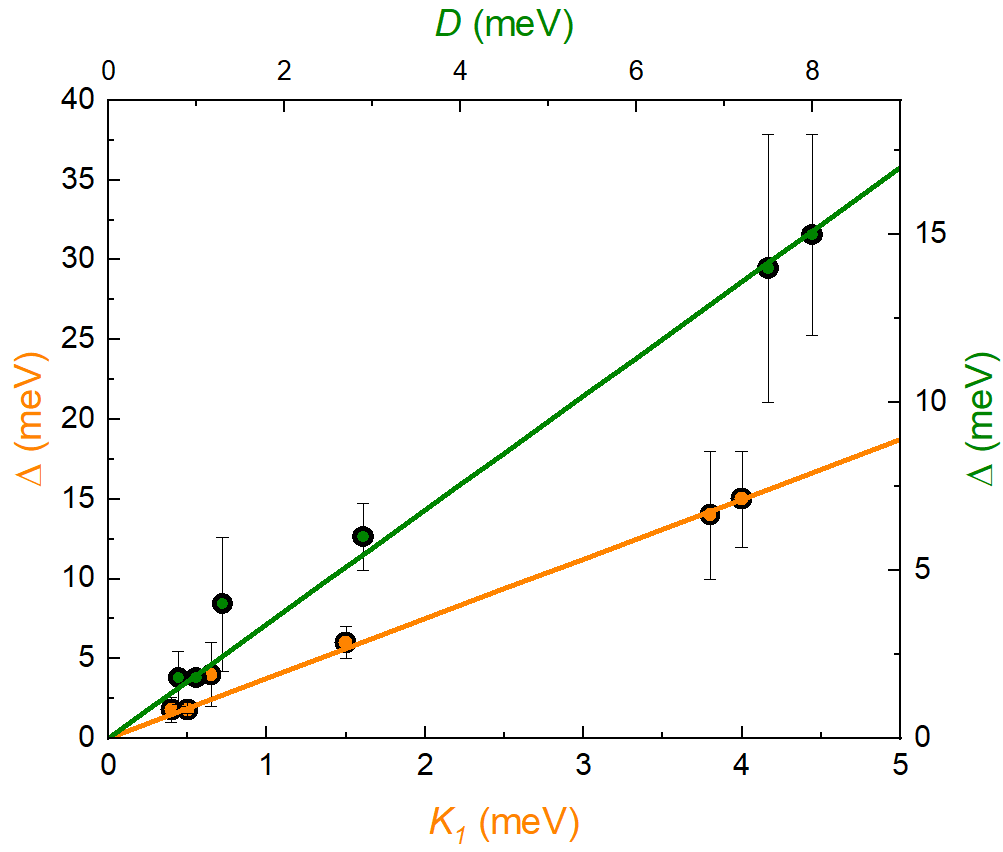}%
\caption{\label{fig:Delta_vs_KD}The variation of the measured spin gap, $\Delta$, with the anisotropic exchange parameter $K_1$ (in orange) and single-ion anisotropy \textit{D} (in green) which are responsible for generating spin gaps within the model Hamiltonians.}
\end{figure}

\section{Discussion}

The gapped magnetic excitation spectrum in the weakly monoclinic, double perovskite family La$_2$LiRuO$_6$ and La$_2$LiOsO$_6$ bears a striking resemblance to that observed in their corresponding cubic counterparts, Ba$_2$YRuO$_6$ and Ba$_2$YOsO$_6$, despite the fact that La$_2$LiRuO$_6$, Ba$_2$YRuO$_6$, and Ba$_2$YOsO$_6$ share a common magnetic structure below their respective ordering temperatures ($T_N$), while La$_2$LiOsO$_6$ displays a different antiferromagnetic structure\cite{Thompson1}. It is therefore useful to make a quantitative comparison between the figures-of-merit for the magnetic properties and energy scales in these two families of double perovskites. This is what is shown in \cref{tab:summary}. We shall also discuss these trends in light of the microscopic spin Hamiltonians we have determined using spin wave theory and shown in \cref{tab:fit_parameters}. 

Table II shows the measured $T_N$, ordered moment size, Curie-Weiss constants ($\theta_{CW}$), spin wave bandwidths, and spin gaps for the cubic double perovskite family Ba$_2$YXO$_6$, and for the monoclinic double perovskite families La$_2$LiXO$_6$ and La$_2$NaXO$_6$, where X = Ru and Os. The ratio of the observed spin gaps within a particular Ru$^{5+}$ and Os$^{5+}$ family is also listed where possible, and, as previously discussed, it is as expected from atomic spin-orbit coupling, $\sim$ 3.4.

One can see that a consistent picture emerges for the energy scales of the monoclinic La$_2$LiRuO$_6$ and La$_2$LiOsO$_6$ family relative to those of the cubic Ba$_2$YXO$_6$ (X = Ru or Os) family, wherein all energy scales in the monoclinic family are smaller than those corresponding to the cubic family by factors of between 2 and 3. Note here that the Curie-Weiss constants for the cubic double perovskites, Ba$_2$YRuO$_6$ and Ba$_2$YOsO$_6$, are well above room temperature, and thus difficult to accurately determine, as the high temperature regime of validity of such an analysis is not easily accessible.  Nonetheless, the conclusion remains that the $\theta_{CW}$ values for Ba$_2$YRuO$_6$ and Ba$_2$YOsO$_6$ are substantially larger than their monoclinic counterparts, La$_2$LiRuO$_6$ and La$_2$LiOsO$_6$.

For this comparison we also include results from related monoclinic double perovskite systems: La$_2$NaRuO$_6$ and La$_2$NaOsO$_6$, and also Sr$_2$ScOsO$_6$ which have all  been recently studied\cite{Aczel1,Aczel2,Taylor1,Taylor2}. To the extent that the crystal structure departure from FCC is quantified by a single crystallographic parameter $\beta$, La$_2$NaRuO$_6$ and La$_2$NaOsO$_6$ display the largest such departure, the largest monoclinicity among these materials. Sr$_2$ScOsO$_6$ displays a monoclinic angle of $\beta =$ 90.22$^{\circ}$ which is larger than that of $\beta =$ 90.15$^{\circ}$ for La$_2$LiOsO$_6$.  It also possesses considerably larger disorder in the form of B/B$^{\prime}$ mixing, ($\sim 5 \%$).  The energy scales of the La$_2$NaRuO$_6$ and La$_2$NaOsO$_6$ family are all suppressed relative to the La$_2$LiXO$_6$ and Ba$_2$YXO$_6$ families; so much so that La$_2$NaOsO$_6$ does not order to temperatures as low as 7 K, while La$_2$NaRuO$_6$ orders into an unusual incommensurate structure below $T_N$ $\sim$ 15 K\cite{Aczel1}. This incommensurate structure is unique among these  $d^3$ systems, and has been attributed to the relatively large monoclinic angle, which tends to introduce considerable tilting of the NaO$_6$ and Os/RuO$_6$ octahedra.  This is argued to randomize the strength of near neighbor exchange interactions, such that they are, on average, weaker.  This could lead to competition between near-neighbor \textit{J}$_1$ interactions and next near-neighbor, \textit{J}$_2$, interactions\cite{Taylor1} leading to incommensurate magnetic structures.  Sr$_2$ScOsO$_6$ is an interesting comparator as it exhibits the highest $T_N$, 92 K, among this grouping, and its spin gap is large (14(4) meV), even though it displays a strong departure from cubic symmetry. Nonetheless it is a well-behaved member of this comparator group, as high $T_N$ and high $\theta_{CW}$ imply a high spin gap and spin wave bandwidth, a similar trend when compared to the other double perovskites in this group.


The wide variation in $T_N$ observed in the $d^3$ systems is considered in Fig. 10, and an excellent linear relationship between $T_N$ and  $S(J_1 + 2K_1)$, going through the origin, is observed.  This is consistent with the observation that $T_N$ scales according to the top of the spin wave band in all these double perovskite magnets.  Figure 11 demonstrates how \textit{K}$_1$ generates the spin gap (as does single-ion anisotropy \textit{D}). \textit{J}$_1$ generates the spin wave bandwidth, while both the bandwidth and the spin gap scale as the moment size, \textit{S}. Hybridization of the $d$ electron orbitals is stronger for the 5$d$ osmates compared to the 4$d$ ruthenates, and this appears to result in a lower ordered moment, \textit{S}, in the osmates compared to the ruthenates. \textit{K}$_1$ is empirically observed to be twice as effective at increasing the gap as \textit{J}$_1$ is to increasing the bandwidth. Both contribute equally to the energy of the top of the spin wave band, and this then gives the relation that $T_N$ is expected to increase as $S(J_1 + 2K_1)$, as Fig. 10 illustrates.  


Table II show that the frustration index $f$, defined as the ratio of $\theta_{CW}$ to $T_N$, is highest for the cubic double perovskites, $\sim$ 10, as expected. The high symmetry of the face centred cubic structure allows the most competititon among equivalent interactions.  This condition is expected to be relaxed somewhat as the symmetry is lowered to monoclinic.  We see that the cubic double perovskites display frustration indices of $\sim$ 10, which is about 30 $\%$ greater than those displayed by the monoclinic double perovskites in this comparator group. A frustration index of 10 is large, comparable to those exhibited, for example, by the 4$d^2$ pyrochlore antiferromagets Y$_2$Mo$_2$O$_7$ and Lu$_2$Mo$_2$O$_7$, both of which exhibit frozen spin glass states at sufficiently low temperatures.  However these $f$ values are not as large as those found in quasi-two dimensional 3$d^9$ Kagome antiferromagnets, such as Herbertsmithite, $\mathrm{ZnCu_3(OH)_6Cl_2}$, where $f$ exceeds 200.  


\begin{table*}[htbp!] 
\caption{\label{tab:summary} A summary of key properties of $d^3$ double perovskites considered in this comparison are shown in this table. Please note that a.) INS studies have not been performed on the Ru based compound, Sr$_2$ScRuO$_6$ and as such $\Delta_{Os}/\Delta_{Ru}$ is not reported for this family of double perovskites, b.) numerical values for $T_N$, $\mu$, $f$ and $\Delta$ are not quoted for La$_2$NaOsO$_6$ as it fails to develop long range order as determined in studies by Aczel \textit{et al}\cite{Aczel1,Aczel2} and c.) the values quoted for $\Delta$ in this table are the values which have been defined using the convention described earlier in Section \ref{SpinW} and will not correspond to the values reported in original work on these materials\cite{Carlo2,Kermarrec,Taylor2,Aczel2}.}
\begin{ruledtabular}\begin{tabular}{cccccccccc}
System & $T_N$ (K) &  $\mu$ & $ \theta_{CW}$ (K) & Bandwidth (meV) & $\beta \left(\degree\right)$ & \textit{f} & $\Delta$ (meV) & $\Delta_{Os}/\Delta_{Ru}$ & Ref. \\ [2ex]
\hline \\
$\mathrm{Ba_2YRuO_6}$ & 36 & 2.2(1)$\mu_B$ & -399(2) & 11(2) & 90 & 11 & 4(2) & \multirow{2}{*}{3.8} & [3]\\
$\mathrm{Ba_2YOsO_6}$ & 69 & 1.65(6)$\mu_B$ & -717(5) & 16(3) & 90 & 11 & 15(3) & & [4] \\
\hline \\
$\mathrm{La_2LiRuO_6}$ & 23.8 & 2.2(2)$\mu_B$ & -185(5) & 5.7(9) & 90.020(5) & 9 & 1.8(8) & \multirow{2}{*}{3.3} & [5,23]\\
$\mathrm{La_2LiOsO_6}$ & 30 &  1.8(2)$\mu_B$ &  -154(2) & 12(1) & 90.147(1) & 6 & 6(1) & & [5]\\
\hline \\
$\mathrm{La_2NaRuO_6}$ & 15(1) & 1.87(7)$\mu_B$ & -57(1) & 2.0(3) & 90.495(2) & 4 & 1.8(2) & \multirow{2}{*}{-} & \multirow{2}{*}{[7,8]} \\
$\mathrm{La_2NaOsO_6}$ & - & - & -74(1) & - & 90.587(2) & - & -  &  & \\ \hline \\
$\mathrm{Sr_2ScOsO_6}$ & 92(1) & 1.6(1)$\mu_B$ & -677 & 20(4) & 90.219(2) & 7.4 & 14(4)  & - & [9,10] \\
\end{tabular}\end{ruledtabular}
\end{table*}

\section{Conclusions}

To conclude, we have studied the inelastic magnetic scattering and corresponding spin gaps in the weakly monoclinic, double perovskite antiferromagnets La$_2$LiRuO$_6$ and La$_2$LiOsO$_6$ using time-of-flight inelastic neutron scattering techniques.  We observe the spin gaps to collapse on passing through $T_N$.  The spin gaps themselves, 1.8(8) meV for La$_2$LiRuO$_6$ and 6(1) meV for La$_2$LiOsO$_6$, scale with the strength of the atomic spin-orbit coupling parameter, $\lambda$, appropriate for 4$d^3$ Ru$^{5+}$ and 5$d^3$ Os$^{5+}$. Such a spin gap is naively unanticipated for a $d^3$ system due to the orbital singlet expected from half-filled t$_{2g}$ levels.  The magnetic spectral weight in La$_2$LiRuO$_6$ and La$_2$LiOsO$_6$ below their respective $T_N$\,s is well described by linear spin wave theory, based on near-neighbor anisotropic exchange, as is the low temperature magnetic spectral weight in the cubic double perovskites Ba$_2$YRuO$_6$ and Ba$_2$YOsO$_6$, which were previously measured.  

A similar spin wave analysis was carried out earlier for Sr$_2$ScOsO$_6$, and the $T_N$s for this family of 5 double perovskite antiferromagnets scales very well with $S(J_1 + 2K_1)$, which characterizes the energy of the top of the spin wave band in all of these materials.  The magnitude of the spin wave gap is controlled by the near-neighbor anisotropic exchange strength, \textit{K}$_1$, and together these are strong evidence for the gapped spectrum arising due to anisotropic exchange, which itself is generated by spin-orbit coupling.  We hope that these new measurements and their analysis in the context of spin dynamics in other $d^3$ double perovskites can guide a full understanding of the nature of their ordered states and counter-intuitive spin gaps.

\section{Acknowledgments}
Research at McMaster University was supported by NSERC of Canada.  This work was supported in part by the National Science Foundation under Grant No. PHYS-1066293 and the hospitality of the Aspen Center for Physics.  We also acknowledge the hospitality of the Telluride Science Research Center. We gratefully acknowledge useful conversations with A. Taylor, R. F. Fishman, and S. Calder. We are very grateful for the instrument and sample environment support provided during our inelastic neutron scattering measurements at SEQUOIA. The experiments which were performed at the Spallation Neutron Source at Oak Ridge National Laboratory was sponsored by the US Department of Energy, Office of the Basic Energy Sciences, Scientific User Facilities Division.

\newpage

\bibliographystyle{apsrev4-1}
\bibliography{d3_Script_14May2018_Final.bib}
 
\end{document}